\documentclass[reprint,aps,prc,amsmath,amssymb,nofootinbib,superscriptaddress]{revtex4-2}

\usepackage[T1]{fontenc}
\usepackage[utf8]{inputenc}
\usepackage{graphicx,color,rotating,pifont}
\usepackage{mathtools}
\usepackage{siunitx}
\usepackage{bm}
\usepackage{ae}

\usepackage{siunitx}
\usepackage{dcolumn}
\usepackage{txfonts}
\usepackage{tensor}
\usepackage{braket}
\usepackage{booktabs}
\usepackage{xspace}
\usepackage{mathrsfs}
\usepackage{amssymb} 
\usepackage{amsmath}
\usepackage{esvect}
\usepackage{diagbox}

\usepackage{hyperref}
\hypersetup{
    colorlinks=true,
    linkcolor=blue,
    filecolor=magenta,      
    urlcolor=blue,
    pdftitle={Overleaf Example},
    pdfpagemode=FullScreen,
    }

\DeclareSIUnit{\fm}{\femto\meter}

\renewcommand{\vec}[1]{\ensuremath{\bm{#1}}}

\newcommand{\beq}{\begin{equation}}
\newcommand{\eeq}{\end{equation}}
\newcommand{\beqn}{\begin{eqnarray}}
\newcommand{\eeqn}{\end{eqnarray}}
\newcommand{\bsub}{\begin{subequations}}
\newcommand{\esub}{\end{subequations}}
\newcommand{\bpm}{\begin{pmatrix}}
\newcommand{\epm}{\end{pmatrix}}

\usepackage{soul,xcolor}
\usepackage[normalem]{ulem}
\setstcolor{red}

\begin{document}
\title{Shell structure and shape transition in odd-$Z$ superheavy nuclei with proton numbers $Z=117, 119$: insights from deformed relativistic Hartree-Bogoliubov in continuum}

\author{Y. X. Zhang}    
\affiliation{School of Physics and Astronomy, Sun Yat-sen University, Zhuhai 519082, P.R. China}  

\author{B. R. Liu}   
\affiliation{School of Physics and Astronomy, Sun Yat-sen University, Zhuhai 519082, P.R. China}   

\author{K. Y. Zhang}   
\affiliation{Institute of Nuclear Physics and Chemistry, China Academy of Engineering Physics, Mianyang, Sichuan 621900, P.R. China}

  \author{J. M. Yao}   
  \email{Corresponding author: yaojm8@sysu.edu.cn}
  \affiliation{School of Physics and Astronomy, Sun Yat-sen University, Zhuhai 519082, P.R. China}

\date{\today}

\begin{abstract}

We present a systematic study on the structural properties of odd-$Z$ superheavy nuclei with proton numbers $Z=117, 119$, and neutron numbers $N$ increasing from $N=170$ to the neutron dripline within the framework of axially deformed relativistic Hartree-Bogoliubov theory in continuum (DRHBc). The results are compared with those of even-even superheavy nuclei with proton numbers $Z=118$ and $120$. We analyze various bulk properties of their ground states, including binding energies, quadrupole deformations, root-mean-square radii, nucleon separation energies, and $\alpha$-decay energies. The coexistence of competing prolate and oblate or spherical shapes leads to abrupt changes in both quadrupole deformations and charge radii as functions of neutron numbers. Compared to even-even nuclei, the odd-mass ones exhibit a more complicated transition picture, in which the quantum numbers of $K^\pi$ of the lowest-energy configuration may change with deformation. This may result in the  change of angular momentum in the ground-state to ground-state $\alpha$-decay and thus quench the decay rate in odd-mass nuclei.  Moreover, our results demonstrate a pronounced proton shell gap at $Z=120$, instead of $Z=114$, which is consistent with the predictions of most covariant density functional theories. Moreover, large neutron shell gaps are found at $N=172$ and $N=258$ in the four isotopic chains, as well as at $N=184$ in the light two isotopic chains with $Z=117$ and $Z=118$, attributed to the nearly-degenerate $3d$ and $4p$ spin-orbit doublet states due to the presence of bubble structure.
\end{abstract}

\maketitle

 \section{Introduction}
 Superheavy nuclei are usually defined as atomic nuclei with proton number $Z \geq 104$ \cite{Hofmann:2000cs}. These transactinide nuclei are unstable due to the strong electrostatic repulsion force between protons and are prone to decay through various radioactive processes, dominated by $\alpha$-decay and spontaneous fissions~\cite{Viola:1966,Mosel:1969ZPhysik}. However, some superheavy nuclei can survive with lifetimes of minutes or even longer due to the large quantum shell effect arising from the motions of neutrons and protons governed by nuclear force~\cite{Myers:1966NP,Sobiczewski:PL1966,Armbruster:1985ARNPS}. Considering this effect in semi-empirical theories, the nucleus with $Z=114$, $N=184$ was predicted to be the next doubly magic  nucleus beyond  \nuclide[208]{Pb}~\cite{Myers:1966NP,Viola:1966,Zhang:2012PRC}.  The study of superheavy nuclei helps us answer the fundamental questions on the limits of the existence of the heaviest elements and the borders of the nuclear chart~\cite{Smits:2023jzt}, as well as understand nucleosynthesis processes in the universe. Therefore, the synthesis of superheavy nuclei and exploration of their structural properties have been at the forefront of nuclear physics~\cite{Hofmann:2000cs,Sobiczewski:2007PPNP,Hamilton:2013ARNPS,Giuliani:2019RMP}.  To date, superheavy nuclei with proton numbers up to $Z=118$ have been synthesized in fusion reactions \cite{Armbruster:1985ARNPS,Oganessian:2006va,Oganessian:PRL2012,Oganessian:2015rpg}, and great efforts are being devoted to the synthesis of elements with $Z=119$ and $Z=120$ \cite{Oganessian:2009zza,Khuyagbaatar:2020PRC,Sakai:2022EPJA,Gan:2022EPJA,Zhou:2022,Zhu:2023}.

The structural properties of superheavy nuclei have been studied with different types of nuclear models~\cite{Nix:1972xx,Bender:2003RMP,Sobiczewski:2007PPNP}, including microscopic-macroscopic models~\cite{Myers:1966NP,Mosel:1969ZPhysik,Nilsson:1969NPA,Patyk:1991NPA,Moller:1992NPA,Wang:2014PLB} and self-consistent mean-field approaches~\cite{Rutz:1997PRC,Bender:1999PRC} based on nonrelativistic~\cite{Cwiok:1996NPA,Cwiok:2005nature} and relativistic~\cite{Lalazissis:1996NPA,Afanasjev:2003gi,Prassa:2012PRC,Li:2014PLB,Agbemava:2015PRC} energy density functionals (EDFs). These methods have even been applied to study hyperheavy nuclei~\cite{Afanasjev:2018PLB,Agbemava:2019PRC,Agbemava:2021PRC}. It is noted that studies on odd-mass and odd-odd superheavy nuclei, especially the neutron-rich ones, are rather scarce, with exceptions including studies within the microscopic-macroscopic method~\cite{Wang:2014PLB,Adamian:2021EPJA,Jachimowicz:2021ADNDT}, the Skyrme Hartree-Fock (HF) plus BCS approach~\cite{Sarriguren:2019PRC}, and the Skyrme Hartree-Fock-Bogoliubov (HFB) approach~\cite{Cwiok:1999PRL}, where the polarization effect due to the odd neutron is treated self-consistently.
   
Shell structure is crucial for the stability of superheavy nuclei, and thus has been explored extensively with different approaches.  Previous studies based on  self-consistent mean-field approaches indicate that the proton and neutron numbers with $(Z=114$, $N=184)$, $(Z=120$, $N=172)$, and $(Z=126$, $N=184)$ are possible magic numbers in superheavy nuclei~\cite{Rutz:1997PRC,Zhang:2005NPA}. However, the specific values vary with the employed parametrizations of EDFs~\cite{Bender:1999PRC}. It has been found that the $Z=114$ and $Z=120$ shell gaps compete with each other. For most Skyrme EDFs, the spin-orbit splitting of the proton $2f$ shell is large, leading to a large $Z=114$ shell gap but a small $Z=120$ shell gap. In contrast, the $Z=114$ gap in the predictions of most relativistic EDFs is slightly smaller, but with a much larger $Z=120$ shell gap. The latter is also attributed to the weak spin-orbit splitting of proton $3p$ doublet states due to the development of a central depression in the nuclear density distribution. On the other hand, it has been demonstrated that shell structure in light and medium-mass nuclei may evolve significantly with neutron number~\cite{Otsuka:2020RMP}, and shell gaps are weakened in neutron-rich nuclei~\cite{Dobaczewski:1994PRL}. In superheavy nuclei, nucleon shell gaps are largely affected by the spin-orbit splittings of neighboring states~\cite{Bender:1999PRC} and nuclear deformations. Therefore, it would be of great interest to systematically study the evolution of shell gaps and shape transition towards neutron dripline in superheavy nuclei.

In this paper, we present a systematic study of odd-$Z$ superheavy nuclei with proton numbers $Z=117$, and  $119$, and neutron numbers $N$ increasing from $N=170$ to the neutron dripline within the framework of an axially deformed relativistic Hartree-Bogoliubov theory in continuum (DRHBc), with emphasis on the evolution of shell structures and shape coexistence.
 In the DRHBc theory, the effects of pairing correlations, deformation, as well as the coupling to continuum states can be treated in a microscopic and self-consistent manner, making it one of the most suitable nuclear models for both stable and neutron-rich deformed nuclei~\cite{Meng:2005PPNP,Zhou:2009sp,Li:2012PRC,Xia:2017ADNDT,Li:2012CPL,Chen:2012PRC}. This method has achieved great success in providing a unified description of a total of 4829 isotopes from the proton dripline to the neutron dripline with $8\leq Z \leq 120$, including even-even nuclei~\cite{Zhang:2020PRC,DRHBcMassTable:2022_EE,Zhang:2021PRC},   even-$Z$, odd-$N$ nuclei~\cite{pan:2022prc,DRHBcMassTable:2024_EO}, and odd-odd nuclei~\cite{He:2024} with proton numbers $Z=106\text{--}112$. This success motivates us to study in detail the structural properties of heavier odd-$Z$ superheavy nuclei with $Z=117, 119$ within the DRHBc theory. The results are compared with those of even-even superheavy nuclei with proton numbers $Z=118$ and $120$.

The article is organized as follows. In Sec.\ref{Sec.model}, we present the main formulas for the DRHBc theory. In Sec.\ref{Sec.result}, we discuss the results of DRHBc calculations for the ground-state properties of $Z=117-120$ superheavy nuclei, including binding energies, quadrupole deformation parameters, potential energy curves, density distributions, root-mean-square (rms) radii, $\alpha$-decay energies, and the evolution of nucleon shell structures with neutron number. Finally, a brief summary and outlook are given in Sec.~\ref{Sec.summary}.

 \section{THEORETICAL FRAMEWORK} 
 \label{Sec.model}
In this work, we employ the DRHBc theory starting from the following effective Lagrangian density in which nucleons interact effectively via the contact interaction of different types of vertices,
 \begin{equation}
 \label{eq:Lagrangian}
 \begin{aligned}
\mathcal{L}&= \bar{\psi}(i\gamma_{\mu}\partial^{\mu}-M)\psi-\frac{1}{2}\alpha_{S}(\bar{\psi}\psi)(\bar{\psi}\psi)-\frac{1}{2}\alpha_{V}(\bar{\psi}\gamma_{\mu}\psi)(\bar{\psi}\gamma^{\mu}\psi)\\
&-\begin{aligned}\frac{1}{2}\alpha_{TV}(\bar{\psi}\vec{\tau}\gamma_{\mu}\psi)(\bar{\psi}\vec{\tau}\gamma^{\mu}\psi)-\frac12\alpha_{TS}(\bar{\psi}\vec{\tau}\psi)(\bar{\psi}\vec{\tau}\psi)-\frac13\beta_{S}(\bar{\psi}\psi)^{3}\end{aligned} \\&-\frac14\gamma_{S}(\bar{\psi}\psi)^{4}-\frac14\gamma_{V}[(\bar{\psi}\gamma_{\mu}\psi)(\bar{\psi}\gamma^{\mu}\psi)]^{2}-\frac12\delta_S\partial_\nu(\bar{\psi}\psi)\partial^\nu(\bar{\psi}\psi)\\
&-\begin{aligned}\frac12\delta_V\partial_\nu(\bar{\psi}\gamma_\mu\psi)\partial^\nu(\bar{\psi}\gamma^\mu\psi)-\frac12\delta_{TV}\partial_\nu(\bar{\psi}\vec{\tau}\gamma_\mu\psi)\partial^\nu(\bar{\psi}\vec{\tau}\gamma_\mu\psi)\end{aligned}\\&-\frac12\delta_{TS}\partial_{\nu}(\bar{\psi}\vec{\tau}\psi)\partial^{\nu}(\bar{\psi}\vec{\tau}\psi)-\frac14F^{\mu\nu}F_{\mu\nu}\\&-e\bar{\psi}\frac{1-\tau_{3}}{2}\gamma^{\mu}\psi A_{\mu},
\end{aligned}
\end{equation}
where $M$ represents the nucleon mass, and $\vec{\tau}$ for the isospin vector. The third component of the isospin vector is $\tau_3=+1(-1)$ for neutrons (protons). The symbols $\alpha_S$, $\alpha_V$, $\alpha_{TS}$, and $\alpha_{TV}$ denote the coupling constants associated with four-fermion contact interaction terms, while $\beta_S$, $\gamma_S$, and $\gamma_V$ represent nonlinear self-interaction terms. Additionally, $\delta_S$, $\delta_V$, $\delta_{TS}$, and $\delta_{TV}$ denote the coupling constants for gradient terms to simulate finite-range effects of nuclear force. Besides, the electromagnetic interaction between protons is described with the electromagnetic field $A_\mu$. The field tensor $F_{\mu\nu}$ of the electromagnetic field is defined as $F_{\mu\nu}=\partial_\mu A_\nu-\partial_\nu A_\mu$.

In the mean-field approximation, one obtains a relativistic EDF from the effective Lagrangian density (\ref{eq:Lagrangian}). A Dirac equation for the single-nucleon wave function is obtained with the variational principle. The pairing correlation between nucleons can be considered by introducing quasiparticles, the wave functions $(U_k, V_k)^T$ of which are determined by the following equation~\cite{Kucharek:1991ZPA,Meng:2005PPNP} 
\begin{equation}
\label{eq:RHB_equation}
\begin{pmatrix}
  h_D-\lambda_{\tau_3}   & \Delta \\
-  \Delta^*          &-  h_D+\lambda_{\tau_3}
\end{pmatrix}
\begin{pmatrix}
U_k\\
V_k\end{pmatrix}
=E_k\begin{pmatrix}U_k\\
V_k\end{pmatrix}
\end{equation}
where $E_k$, $\lambda_{\tau_3}$, and $\Delta$ are  the energy of the $k$-quasiparticle state, Fermi energy, pairing potential, respectively.  In the even-even nuclear system with time-reversal symmetry, the single-particle Hamiltonian $h_D$ has the following form,
\begin{equation}
 h_D(\boldsymbol{r})
 =\boldsymbol{\alpha}\cdot\boldsymbol{p}+V(\boldsymbol{r})+\beta(M+S(\boldsymbol{r})),
 \end{equation}
where $S(\boldsymbol{r})$ and $V(\boldsymbol{r})$ denote scalar and vector potentials, 
\beqn 
S(\boldsymbol{r})~&=&~\alpha_S\rho_S+\beta_S\rho_S^2+\gamma_S\rho_S^3+\delta_S\Delta\rho_S, \\
V(\boldsymbol{r})~&=&~\alpha_V\rho_V+\gamma_V\rho_V^3+\delta_V\Delta\rho_V+e\frac{1-\tau_3}{2}A_0\nonumber\\
&&+\alpha_{TV}\tau_3\rho_{TV}+\delta_{TV}\tau_3\Delta\rho_{TV}, 
\eeqn 
with the scalar, vector and isovector densities defined as below,
\begin{equation}
\begin{aligned}\rho_{S}(\boldsymbol{r}) &=\sum_{k>0}V_k^\dagger(\boldsymbol{r})\gamma_0V_k(\boldsymbol{r}), \\
\rho_{V}(\boldsymbol{r}) &=\sum_{k>0}V_k^\dagger(\boldsymbol{r})V_k(\boldsymbol{r}), \\
\rho_{TV}(\boldsymbol{r}) &=\sum_{k>0}V_k^\dagger(\boldsymbol{r})\tau_3V_k(\boldsymbol{r}). 
\end{aligned}
\end{equation}
Here the no-sea approximation is adopted, i.e., the summation
runs over the quasiparticle states in the Fermi sea only.

In order to obtain energy curve as a function of quadrupole deformation parameter $\beta^{(t)}_{20}$, we add a constraint on the mass quadrupole operator $\hat Q_{20}\equiv  r^2 Y_{20}$, in which case, the single-particle Hamiltonian is replaced by
\beq 
  h_D - C_0 \Big(\bra{\Phi(\beta^{(t)}_{20})}\hat Q_{20} \ket{\Phi(\beta^{(t)}_{20})} -q_{20}\Big) \hat Q_{20}.
\eeq 
Here, $\ket{\Phi(\beta^{(t)}_{20})}$ represents the mean-field wave function, $q_{20}$ denotes the quadrupole moment, and $C_0$ stands for the corresponding stiffness constant. The axial quadrupole deformation $\beta^{(t)}_{20}$ (dimensionless) of the total nucleus is defined by
\begin{equation}\label{1-1}
\beta^{(t)}_{20}=  \frac{4 \pi}{3 R^2 A}\bra{\Phi(\beta^{(t)}_{20})}\hat Q_{20} \ket{\Phi(\beta^{(t)}_{20})},
\end{equation}
where $R=1.2 A^{1 / 3} \mathrm{fm}$,  and $A$ is the mass number.

By employing  a density-dependent $\delta$  force for the isovector pairing correlation, the pairing potential in the coupled spin $S=0$ and isospin $T=1$ channel is simplified as \cite{Li:2012PRC}
\beq 
 \Delta(\bm{r})
= V_0  \left(1-\frac{\rho_V(\boldsymbol{r})}{\rho_{\rm sat}}\right)  \kappa(\boldsymbol{r}),
\eeq 
with the saturation density $\rho_{\rm sat}=0.152$ fm$^{-3}$ and the pairing tensor 
\begin{equation}
\kappa(\boldsymbol{r})=\sum_{k>0} V_k^\dagger(\boldsymbol{r})U_k(\boldsymbol{r}).
\end{equation}
The total energy of an even-even nucleus is given by~\cite{Meng:1998NPA,Meng:2005PPNP}
\begin{equation}
\label{eq:total_energy}
\begin{aligned}
E&=\sum_{k>0}(\lambda-E_k)\int d^3\boldsymbol{r}V_k^\dagger(\boldsymbol{r})V_k(\boldsymbol{r})\\
&-\int d^3\boldsymbol{r}\Bigg(\frac12\alpha_S\rho_S^2+\frac12\alpha_V\rho_V^2+\frac12\alpha_{TV}\rho^2_{TV} \\
&+\frac23\beta_{S}\rho_{S}^{3}+\frac34\gamma_{S}\rho_{S}^{4}+\frac34\gamma_{V}\rho_{V}^{4}+\frac12\delta_{S}\rho_{S}\Delta\rho_{S} \\
& +\frac12\delta_{V}\rho_{V}\Delta\rho_{V}+\frac12\delta_{TV}\rho_{TV}\Delta\rho_{TV}+\frac12eA_{0}\rho^{(p)}_{V}\Bigg)\\
&-E_\text{pair}+E_{\mathrm{c.m.}},
\end{aligned}
\end{equation} 
where the pairing energy is
\begin{equation}
E_{\mathrm{pair}}=-\frac12\int d^3\boldsymbol{r}\kappa(\boldsymbol{r})\Delta(\boldsymbol{r}),
\end{equation}
and the correction from the center-of-mass motion to the energy 
\begin{equation}
E_{\mathrm{c.m.}}=-\frac1{2AM}\langle\hat{\boldsymbol{P}}^2\rangle. 
\end{equation}
Here, $\hat{\boldsymbol{P}}$ is the total momentum for the nucleus. 

For odd-$Z$ superheavy nuclei, the blocking effect from the unpaired proton is considered by exchanging the $k_b$-th quasiparticle wavefunction $(U_{k_b}, V_{k_b})$ with $(V_{k_b}^*, U_{k_b}^*)$, and its quasiparticle energy $E_{k_b}$ is replaced by $-E_{k_b}$~\cite{ring2004nuclear,Li:2012xaa}. By employing equal-filling approximation (EFA),  the time-reversal invariance is retained, and all spatial components of currents vanish. In this case, the densities and pairing tensors should be replaced as follows~\cite{Li:2012CPL}
\bsub
\beqn 
 \rho^{\prime}_V &=&\rho_V +\frac{1}{2}\left(U_{k_b} U_{k_b}^{* T}-V_{k_b}^* V_{k_b}^T\right), \\
 \kappa^{\prime} &=&\kappa-\frac{1}{2}\left(U_{k_b} V_{k_b}^{* T}+V_{k_b}^* U_{k_b}^T\right).
\eeqn
\esub 
The total energy for an odd-mass nucleus has the same expression at that in Eq. (\ref{eq:total_energy}), provided that the first term (written in the canonical basis) is replaced by~\cite{pan:2022prc}
\beqn 
  2 \sum_{k>0} \left(\lambda-E_k\right) v_k^2+\left(\lambda+E_{k_b}\right) u_{k_b}^2   -\left(\lambda-E_{k_b}\right) v_{k_b}^2,
\eeqn

The rms radii of neutrons and protons are calculated by
\begin{equation} 
R_{\tau_3}=\sqrt{\frac1{N_{\tau_3}}\int d^3\boldsymbol{r}r^2\rho^{(\tau_3)}_{V}(\boldsymbol{r})},
\end{equation}
where $N_{\tau_3}$ is the number of neutrons or protons, and $\rho^{(\tau_3)}_{V}(\boldsymbol{r})$ the corresponding vector density.  The densities $\rho_i$, pairing tensor $\kappa$, and potentials are  expanded in terms of the Legendre polynomials for axially symmetric and spatial
reflection symmetric nuclei,
\begin{equation}
\label{eq:multipole_expansion}
f(\boldsymbol{r})=\sum_{L\geq 0} f_L(r)P_L(\cos\theta),
\end{equation}
with $L$ taking the even number truncated up to $L_{\rm max}$. The radial function is given by
\begin{equation}
f_L(r)=\frac{2L+1}{4\pi}\int d\Omega f(\boldsymbol{r})P_L(\cos\theta).
\end{equation}
 The rms charge radius is calculated from the proton radius,
\begin{equation}
R_\mathrm{ch}=\sqrt{R_p^2+0.64\mathrm{~fm}^2}.
\end{equation}
The correction from the restoration of rotational symmetry to the energy is considered as follows,
\begin{equation}
\label{eq:rotation_correction_E}
E_\mathrm{rot}=-\frac{\langle\hat{\boldsymbol{J}}^2\rangle}{2\mathcal{I}_{\mathrm{IB}}},
\end{equation}
where $\hat{\boldsymbol{J}}$ is the total angular momentum of the nucleus, and $\mathcal{I}_{\mathrm{IB}}$ is the moment of inertia calculated using the Inglis-Belyaev formula~\cite{ring2004nuclear}.

 \section{Results and discussion}
 \label{Sec.result}
 
In this work, we adopt the PC-PK1~\cite{Zhao:2010hi} for the  parameters in the Lagrangian density (\ref{eq:Lagrangian}). The pairing strength is chosen as $V_0=-325$ MeV fm$^{3}$ for both neutrons and protons. The equation (\ref{eq:RHB_equation}) for quasiparticle wave functions ($U_k, V_k$) is solved by expanding them in a spherical Dirac Woods-Saxon basis~\cite{Zhou:2003PRC,Zhang:2022PRC}.  The angular momentum cutoff $J_{\rm max}=23\hbar/2$, the energy cutoff $E_{\rm cut}^{+}=300$ MeV, the box size $R_{\rm max }= 20$ fm, the Legendre expansion truncation $L_{\rm max}= 10$, and the mesh size $\Delta r= 0.1$ fm are taken for the Dirac Woods-Saxon basis.    
  
 \subsection{Bulk properties}
  \label{Sec:Ebinding_deformation}
  
\begin{figure}[] 
\includegraphics[width=\columnwidth]{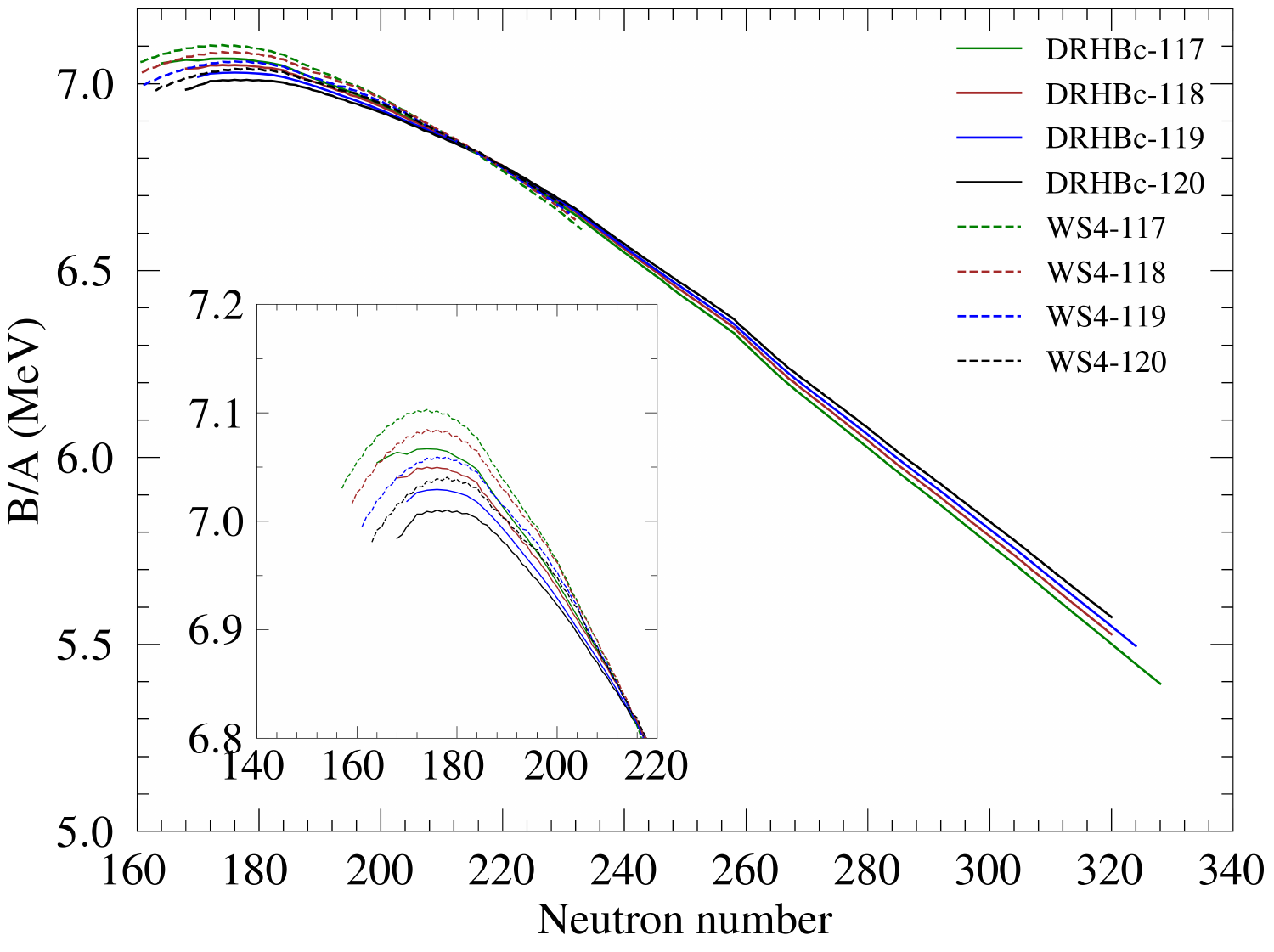}   
\caption{(Color online) The average binding energy per nucleon $B/A$ (MeV) of the ground states of $Z=117-120$ isotopes as a function of the neutron number from the DRHBc calculations using the PC-PK1 EDF, in comparison with the WS4 mass model~\cite{Wang:2014PLB}. In the DRHBc results, both odd-$N$ and even-$N$  nuclei are considered for $Z=118, 120$ isotopic chains, while only even-$N$ nuclei are considered for $Z=117, 119$ isotopic chains.}
\label{fig:energy_per_nucleon} 
\end{figure}

\begin{figure}[] 
\includegraphics[width=\columnwidth]{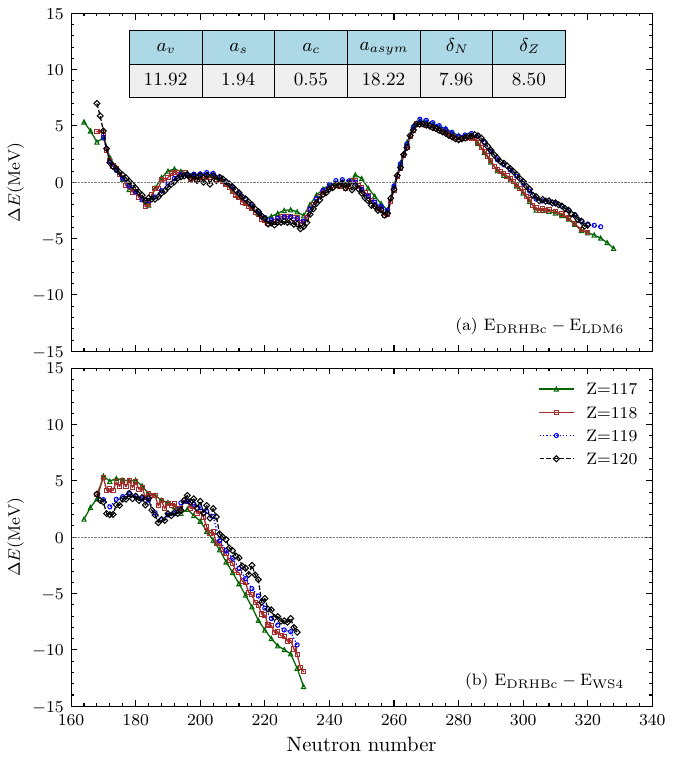}   
\caption{(Color online) The discrepancy of the total energies of the ground states of $Z=117-120$ isotopes, (a) between the DRHBc and LDM6 model, and (b) between the DRHBc and WS4 mass model. }
\label{fig:discrepancy_in_B} 
\end{figure}

\begin{figure}[] 
\includegraphics[width=\columnwidth]{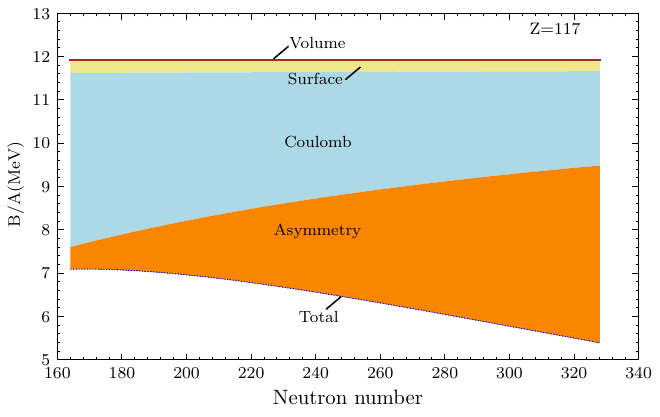}   
\caption{(Color online) The various terms (\ref{eq:LDM6}) contributing to the average binding energies of $Z=117$ isotopes in the LDM6 model, as a function of neutron number, where the total average binding energy is given by the cancellation of the volume term with other terms. The pairing term (about 0.04 MeV per nucleon) is too small to be seen in the figure. See text for details.}
\label{fig:LDM6} 
\end{figure}

Figure~\ref{fig:energy_per_nucleon} displays the binding energies per nucleon for the ground states of the four isotopic chains with $Z=117\text{--}120$ as a function of neutron number. The ground state of each isotope is determined by the lowest-energy state of the DRHBc calculations starting from 11 different initial states with the quadrupole deformation parameter $\beta^{(t)}_{20}$ ranging from $-0.4$ to $+0.6$ and a step size of $\Delta\beta^{(t)}_{20}=0.1$.  For each odd-mass nucleus, the so-called “automatic blocking” procedure~\cite{pan:2022prc} is employed to find out the lowest-energy state. Since there are no  experimental data on these isotopes, we compare our results with the WS4 mass model~\cite{Wang:2014PLB}. It is shown that the $B/A$ of each isotopic chain has a peak at either $N=174$ or $176$. Beyond this nucleus, it decreases smoothly with the increase of neutron number. These features are observed in the results of both the DRHBc theory and WS4 model. The systematic behavior of $B/A$ is mainly governed by  the cancellation between the volume contribution and those from Coulomb energy and asymmetry energy. To see it more clearly, we carry out a linear regression for the total binding energies of totally 460 isotopes by the DRHBc theory based on the formula of a liquid drop model (LDM) with six unknown coefficients,
\beqn
\label{eq:LDM6}
B_{\rm LDM6}(A,Z)
&=&a_V A-a_S A^{2/3}-a_C\frac{Z(Z-1)}{A^{1/3}}-a_{\rm asym}\frac{(A-2Z)^2}A\nonumber\\
&&+\frac{\delta_N[1+(-1)^N]+\delta_Z[1+(-1)^Z]}{A^{1/2}},
\eeqn
where the values of the coefficients $a_i$ and $\delta_{N,Z}$ are given in Fig.~\ref{fig:discrepancy_in_B}(a). The results of the linear regression are labeled as "LDM6" with the rms error of $2.63$ MeV. The discrepancies between the total energies of the superheavy nuclei by the LDM6 and DRHBc are plotted in Fig.~\ref{fig:discrepancy_in_B}(a). Those  between the total energies by the  WS4 and DRHBc  are plotted in Fig.~\ref{fig:discrepancy_in_B}(b) for comparison. It is observed that the energy difference between the DRHBc and LDM6 is oscillating  with the neutron number in the four isotopic chains, attributed to the missing shell effects in the LDM6. In contrast, the discrepancy in the total energies of nuclei by the DRHBc and WS4 is monotonically decreasing with the neutron number. In other words, the neutron-rich superheavy nuclei by the DRHBc are more and more bound  towards neutron dripline, compared to the WS4 model. Since Coulomb energy increases quadratically with proton number, heavy and superheavy nuclei tend to have more neutrons than protons for a given mass number. The most tightly bound superheavy nuclei, corresponding to the peak position in Fig.~\ref{fig:energy_per_nucleon} for a given proton number $Z$, fulfilling approximately the relation\footnote{The contribution from the surface term is dropped out. This term is minor in the LMD6 model for the superheavy nuclei of concerned.}
\beq 
\frac{ZA^{2/3}}{(N-Z)}\simeq 3\frac{a_{\rm asym}}{a_C},
\eeq 
are thus determined by the balance between Coulomb energy and asymmetry energy. Moreover, it is clearly shown in Fig.~\ref{fig:LDM6}  that the decrease of the Coulomb energy (due to the increase of proton radii) and the increase of the asymmetry energy are responsible for the decrease of the average binding energy with the increase of neutron number in each isotopic chain, cf. Fig.~\ref{fig:energy_per_nucleon}.

\begin{figure}[] 
\includegraphics[width=\columnwidth]{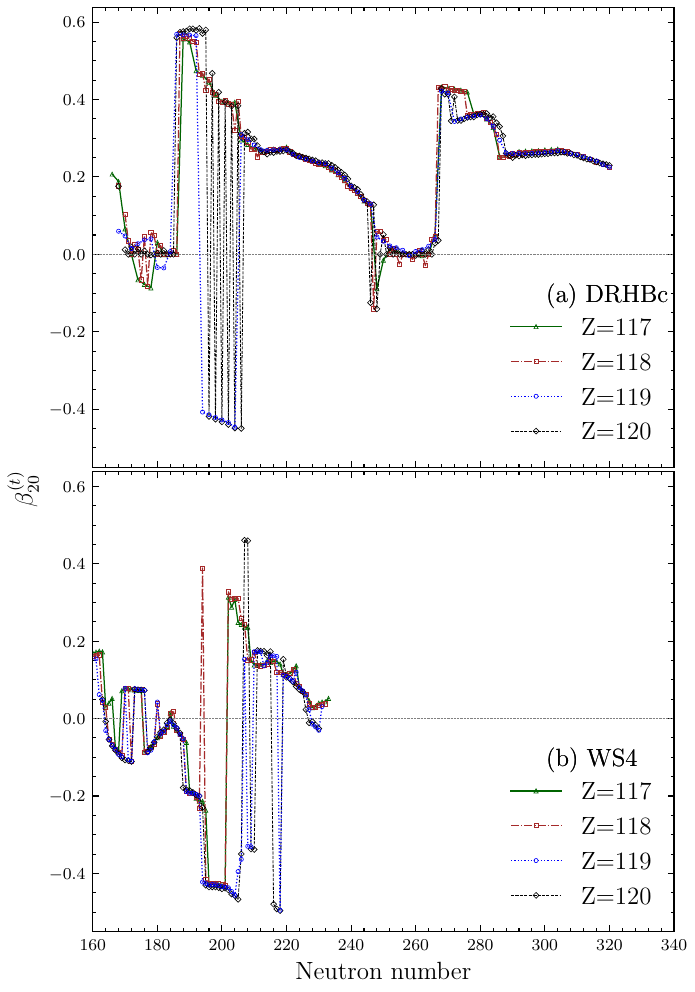}   
\caption{(Color online) The quadruple deformation parameters $\beta^{(t)}_{20}$ of the energy-minimal states in $Z=117-120$ isotopes, as a function of the neutron number, from (a) the DRHBc calculation, in comparison with (b) the WS4 mass model.}
\label{figs:beta2_comparison} 
\end{figure}

\begin{figure}[] 
\includegraphics[width=0.85\columnwidth]{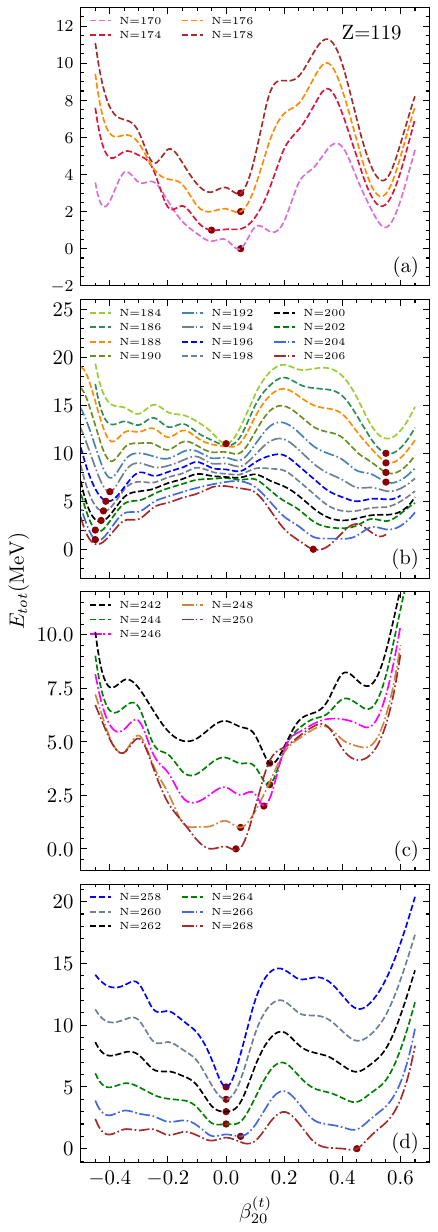}   
\caption{(Color online) The total energies of states in $Z=119$ isotopes as a function of the intrinsic quadrupole deformation parameter $\beta^{(t)}_{20}$, where the neutron numbers are (a) $N=170, 174, 176, 178$, (b) $N=184-206$, (c) $N=242-250$, and (d) $N=258-268$,  respectively.  All energies are normalized to their ground states (indicated with bullets), but with an additional energy shift of 1 MeV between two neighboring isotopes.  }
\label{figs:PECs_Z119} 
\end{figure}

Figure~\ref{figs:beta2_comparison}(a) shows the variation of quadrupole deformation of the ground states in the $Z=117\text{--}120$ isotopes from the DRHBc calculation as a function of neutron number. The results of calculations in the WS4 model are plotted in Fig.~\ref{figs:beta2_comparison}(b) for comparison.  It is shown that the ground states of the four isotopic chains with $N<186$ are spherical or weakly deformed. With the increase  of neutron number, the ground state is shifted to the large prolate deformed state with $\beta_2\simeq 0.6$ which corresponds to the second energy minimum in the light isotopes with $N<186$. When the neutron number $N$ increases up to about $194$, the ground states of isotopes with $Z=119, 120$ are shifted to the oblate state with $\beta_2\simeq -0.4$. A
similar shape evolution has been found, for instance, in the isotopes with $Z = 52 \text{--}56$~\cite{Guo:2023PRC}.  When the neutron number increases up to $N=206$, the ground state is shifted back to the prolate deformed state with $\beta\simeq 0.3$. In contrast, the $Z=117, 118$ isotopes remain prolate states.  When increasing further the neutron number up to $N=248$, the quadrupole deformations of the ground states for all the four isotopic chains decrease smoothly to zero. The systematical behavior of $Z=119$ isotopic chain is clearly shown in  Fig.~\ref{figs:PECs_Z119}. The evolution of the quadrupole deformation of the ground state with neutron number in the WS4 mass model is somewhat different from that of the DRHBc theory. In the WS4 model, the quadrupole deformation decreases from zero to $-0.5$ as neutron number increases from 184 to 204 in the four isotopic chains. In the isotopes with $N\simeq 200$, the WS4 model predicts all the nuclei in the four isotopic chains to have oblate deformed shapes. When $N>208$, the isotopes are mainly prolate deformed, with the quadruple deformation decrease from 0.2 to zero. Generally, the evolution behaviors of the quadrupole deformations predicted by the two methods are similar, but quantitatively different, which might be correlated to the observed discrepancy in their predicted binding energies, cf. Fig.~\ref{fig:discrepancy_in_B}(b). Since these two methods are significantly different, many factors may contribute to the discrepancy and needs further studies. The investigation of the origin of this discrepancy is beyond the scope of this work.

\begin{figure}[] 
\includegraphics[width=0.4\paperwidth]{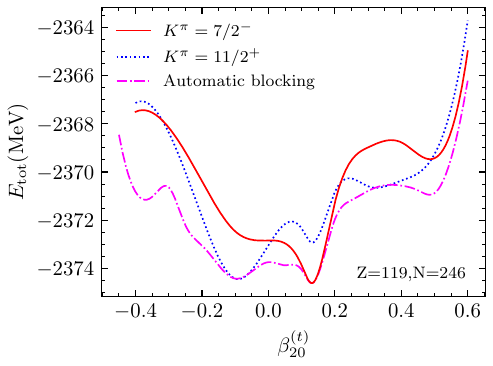}   
\caption{(Color online) The total energies of states with $K^\pi=7/2^-$ (red solid line) and $11/2^+$ (blue dotted line) in $^{365}119$ as a function of the intrinsic quadrupole deformation parameter $\beta^{(t)}_{20}$. The energies of the lowest-energy states (magenta dashed-dotted line) at each quadrupole deformation are also plotted for comparison.}
\label{figs:PES_Z119_N246} 
\end{figure}

\begin{figure}[] 
\includegraphics[width=0.4\paperwidth]{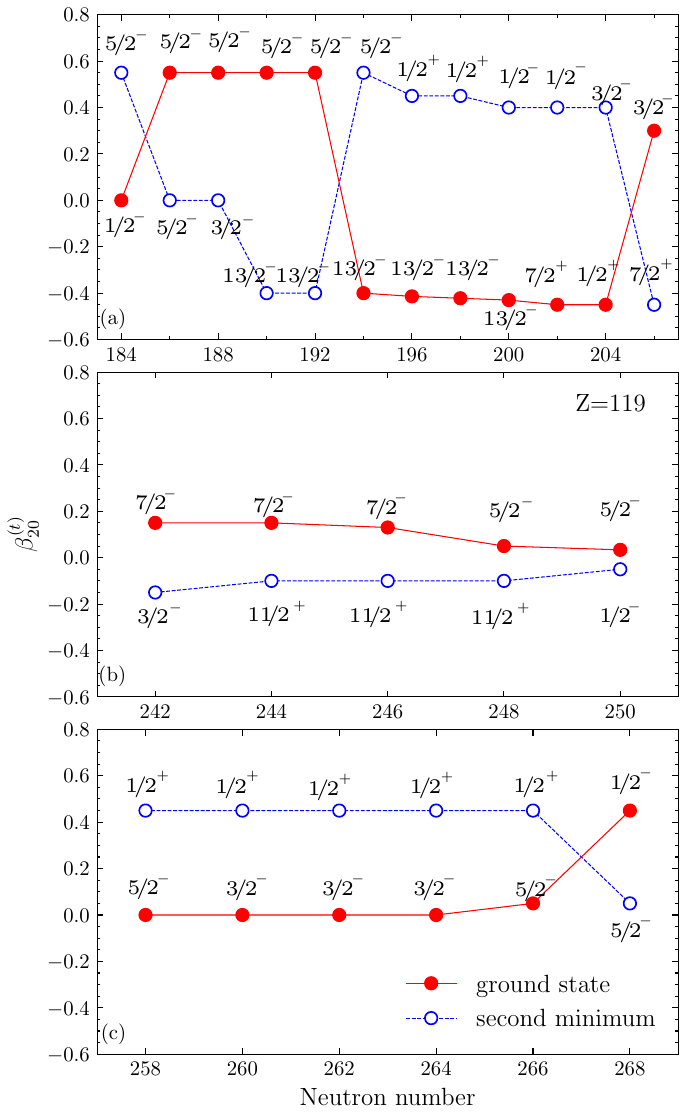}   
  \caption{(Color online) The quadrupole deformation parameters $\beta^{(t)}_{20}$ of the first two lowest-energy minima states (indicated with filled red circles and open blue circles, respectively) in $Z=119$ isotopes from the DRHBc calculations with PC-PK1, where the neutron numbers are (a) $N=184\text{--}206$, (b) $N=242\text{--}250$, and (c) $N=258\text{--}268$, respectively. }    
  \label{figs:Z119_beta2_Kpi}     
\end{figure}

The energy curve for each isotope in Fig.~\ref{figs:PECs_Z119} corresponds the lowest-energy state from the automatic blocking calculation. Different energy-minimal states in the energy surface may have different quantum numbers $K^\pi$, where $K$ is the 3rd-component of the total angular momentum and $\pi$ is the parity of the state. For axially deformed mean-field states, $K^\pi$ are conserved. We take the nucleus with $Z=119, N=246$ as an example to illustrate this point, where the energies of states with $K^\pi=7/2^-, 11/2^+$ are plotted as a function of the quadrupole deformation in  Fig.~\ref{figs:PES_Z119_N246}. It is observed that the global energy-minimum state is found for the configuration with $K^\pi=7/2^-$ at $\beta_{20}\simeq 0.15$. The second energy minimum is located at $\beta_{20}\simeq -0.10$ with  $K^\pi=11/2^+$ whose energy is slightly higher than that of the global minimum.  Figure~\ref{figs:Z119_beta2_Kpi} displays the changes in the quadrupole deformations and quantum numbers $K^\pi$ of the two lowest-energy-minimum states in the $Z=119$ isotopes. One can clearly observe the exchange of configurations of these two states in the nuclei with $N=194$ and $N=268$, where the $K^\pi$ of the global energy minimum state is switched from $5/2^-$ to $13/2^-$ and from $5/2^-$ to $1/2^-$, respectively. This may has a significant impact on their $\alpha$-decay properties which will be discussed in detail later. It illustrates the complexity and challenge in describing the low-energy structure of odd-mass nuclei~\cite{Schunck:2023prc}.

\begin{figure}[] 
\includegraphics[width=0.4\paperwidth]{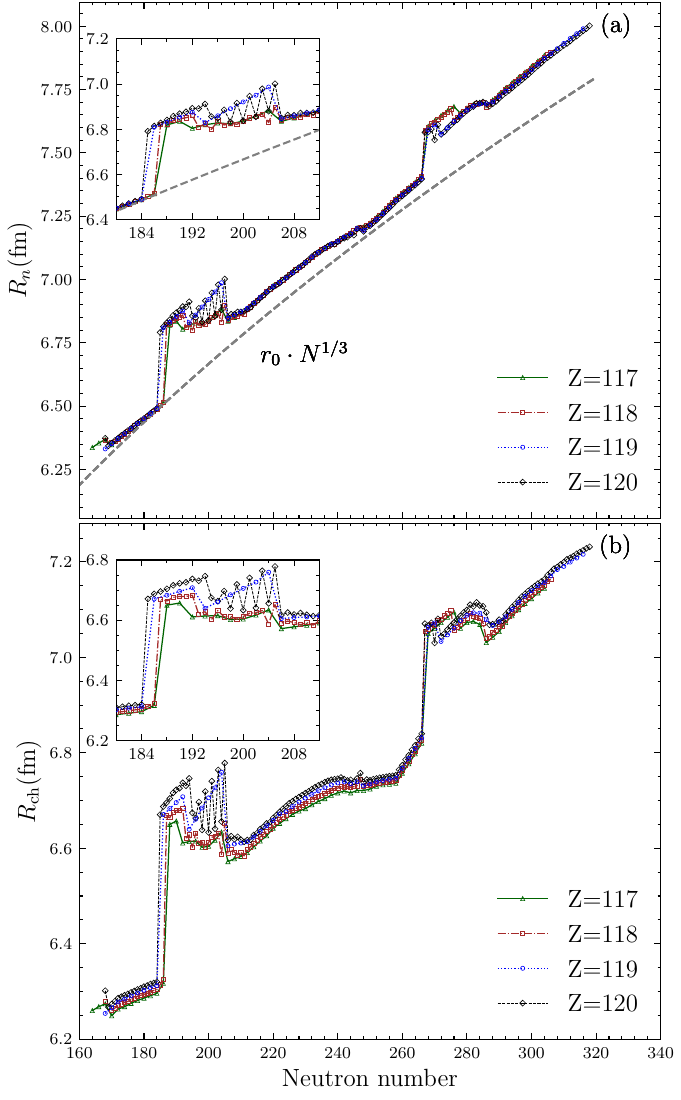}   
\caption{(Color online) (a)   The rms radii $R_n$ of neutrons and (b) charge radii $R_{\rm ch}$ in the nuclei of the $Z=117-120$ isotopic chains as a function of neutron number. The empirical formula $R_n=1.141N^{1/3}$~\cite{DRHBcMassTable:2024_EO} is also given for comparison.}
\label{figs:radii_evolution_N} 
\end{figure}

Both neutron and charge radii are sensitive to the shape of ground state. Figure \ref{figs:radii_evolution_N} shows the variations of neutron rms radius $R_n$ and charge radius $R_{\rm ch}$ of the ground state as a function of neutron number. The empirical formula $R_n=r_0N^{1/3}$ with $r_0=1.140$ fm~\cite{DRHBcMassTable:2024_EO} is plotted for comparison. It is shown that the global behavior of the neutron radii follows the empirical formula. However, a sudden increase of both neutron and charge radii in the ground state around $N=186$, followed by a decrease at $N=206$ are found in all the four isotopic chains. The onset of large radii in these isotopes is related to the shape transition from spherical or weakly deformed shapes to large prolate or oblate ones as seen in Figs.~\ref{figs:beta2_comparison} and \ref{figs:PECs_Z119}. It is shown that the second energy minimum with $\beta_{20}\simeq 0.55$ in $^{289}119$ is higher than the global energy minimum with $\beta_{20}=0.05$ by 1.15 MeV. When the neutron number $N$ increases up to $N=186$, the prolate energy minimum state becomes the ground state, responsible for the sudden increase of radii. A similar phenomenon is observed in the isotopes with neutron number between $N=268$ and $N=286$, attributed to the onset of large prolate deformed shape with quadrupole deformation $\beta_2\simeq 0.3\text{--}0.4$. Quantitatively, the charge radii of the four isotopic chains are slightly different, i.e.,  the nuclei with $Z=119, 120$ and $N\simeq 200$ possess charge radii approximately $0.1$ fm greater than those of their isotones with $Z=117$ and $118$, where quadrupole deformation also contributes.

\subsection{Shell structure and spin-orbit interaction}

\begin{figure}[] 
\includegraphics[width=1.1\columnwidth]{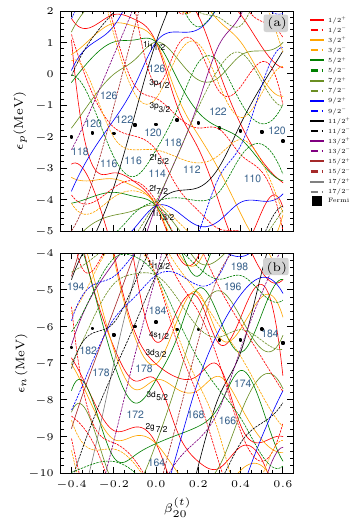}   
\caption{(Color online) Nilsson diagram  for (a) protons and (b) neutrons in $^{304}120$ as a function of the axial deformation parameter $\beta^{(t)}_{20}$. All single-particle energy levels are labeled with $K^\pi$, where positive parity states are represented by solid lines and negative parity with dashed lines. The Fermi energies are indicated with black filled square.   }
\label{figs:Nilsson_diagram_Z120} 
\end{figure}

\begin{figure}[] 
\includegraphics[width=\columnwidth]{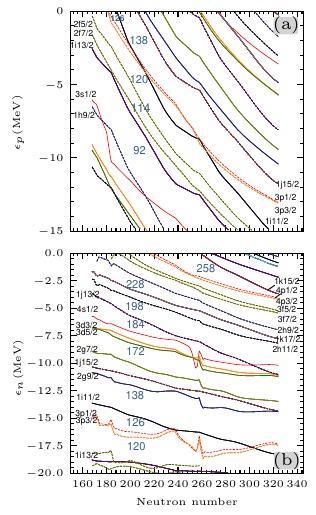}   
\caption{(Color online) The single-particle energy levels of (a) protons and (b) neutrons in the spherical states of $Z=120$ isotopes as a function of neutron number. An evident discontinuity occurs at $N=258$, where pairing correlation between neutrons collapses. }
\label{fig:spherical_shell_structure_evolution_N} 
\end{figure}

\begin{figure}[] 
\includegraphics[width=\columnwidth]{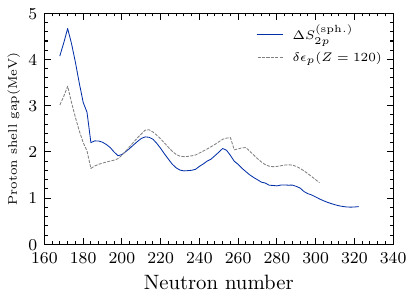}   
\caption{(Color online) Comparison of the two-proton shell gap  $\Delta S^{(\rm sph.)}_{2p}$ (blue solid lines) and the energy gap $\delta \epsilon_p$ (gray dashed lines) between two proton  energy levels around $Z=120$ as a function of neutron number.}
\label{fig:two-proton-shell-gaps_Z120} 
\end{figure}

\begin{figure}[] 
\includegraphics[width=0.9\columnwidth]{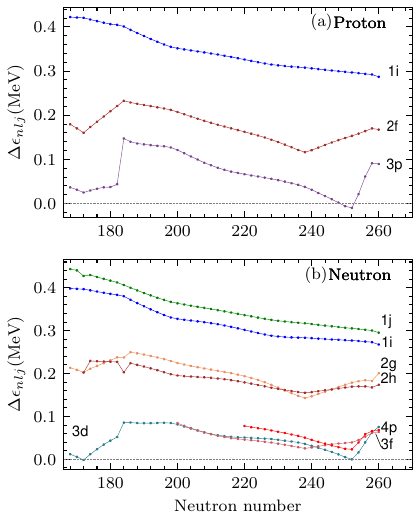}   
\caption{(Color online) The spin-orbit splitting of (a) neutrons  and (b) protons in the spherical states of $Z=120$ isotopes as a function of neutron number.}
\label{fig:spin_orbital_splitting_N} 
\end{figure}

\begin{figure}[] 
\includegraphics[width=\columnwidth]{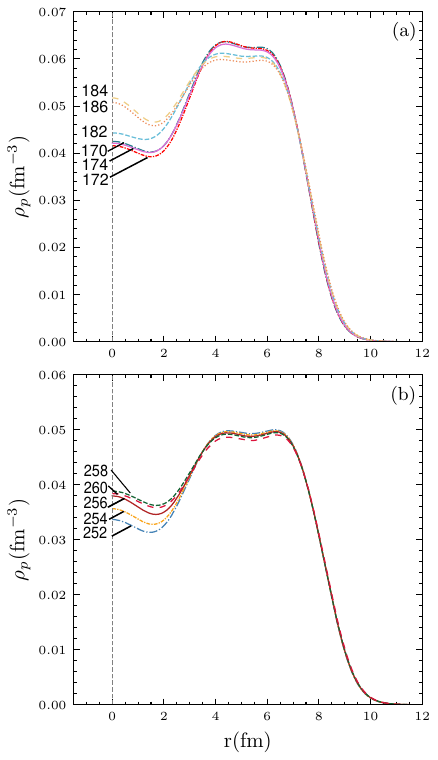}   
\caption{(Color online) The $L=0$ component (\ref{eq:multipole_expansion}) of the proton density in  $Z=120$ isotopes with neutron number (a)  $N=170-186$ and (b) $N=252-260$, respectively.}
\label{fig:density_distribution_proton} 
\end{figure}

\begin{figure}[] 
\includegraphics[width=\columnwidth]{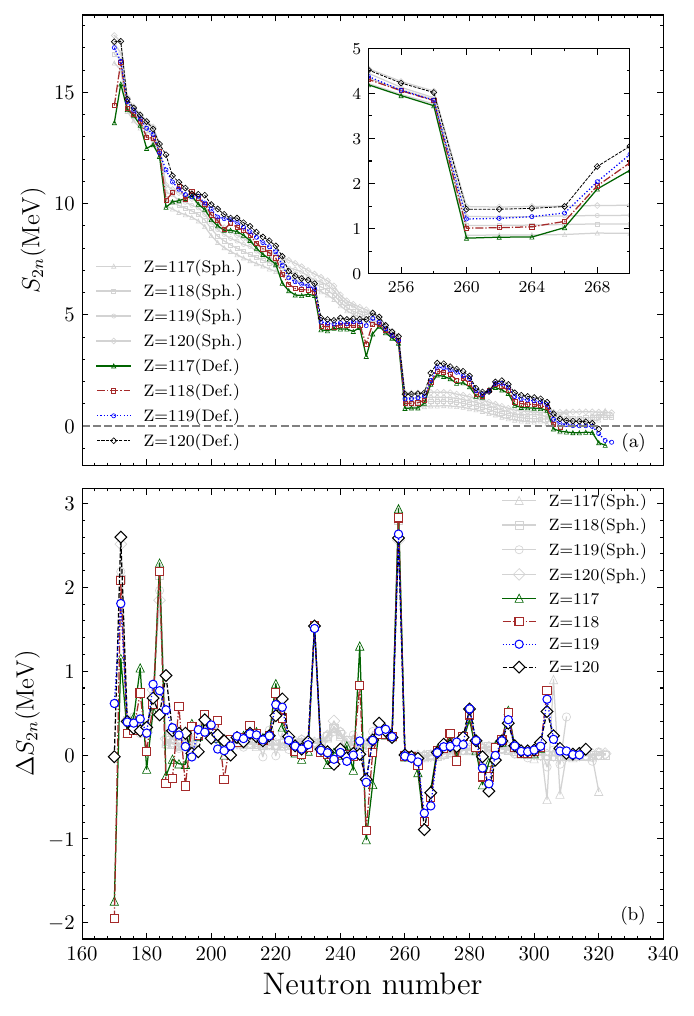}   
\caption{(Color online) (a) The two-neutron separation energies  $S_{2n}$ and (b) 
 their differentiates $\Delta S_{2n}$ in the $Z=117-120$ isotopic chains as a function of neutron number. The results of calculations for the states restricted to have the spherical shape are also plotted for comparison.}  
\label{figs:s2ndelta2n} 
\end{figure}

\begin{figure}[] 
\includegraphics[width=\columnwidth]{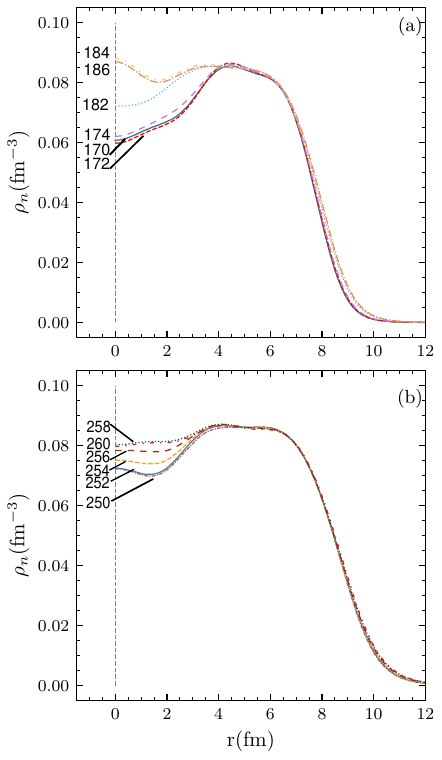}   
\caption{(Color online)  Same as Fig.~\ref{fig:density_distribution_proton}, but for neutrons.}
\label{fig:density_distribution_neutron} 
\end{figure}

\begin{figure*}[] 
\includegraphics[width=0.7\paperwidth]{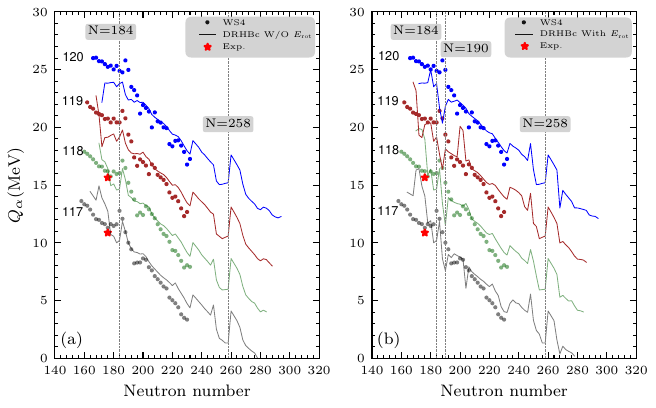} 
\caption{(Color online) The $Q_\alpha$ values of the four isotopic chains with $Z=117-120$ as a function of neutron number from the DRHBc calculations, in comparison with the results of WS4 model. The values of two neighbouring isotopic chains are shifted by 4 MeV. The isotopes with only even neutron numbers are considered. (a) and (b) shows the results of DRHBc calculations without and with the energy (\ref{eq:rotation_correction_E}) from the restoration of rotational symmetry, respectively. The data are taken from Refs.~\cite{Oganessian:2006va,Oganessian:PRL2012}.}
\label{fig:Qalpha} 
\end{figure*}

\begin{figure}[] 
\includegraphics[width=\columnwidth]{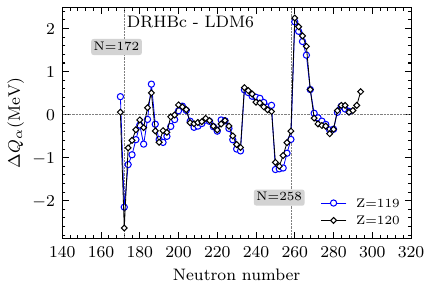} 
\caption{(Color online) The deviation $\Delta Q_\alpha$ of the $\alpha$-decay energies by the LDM6 model from the DRHBc theory (w/o $E_{\rm rot}$) for  the  $Z= 119, 120$ isotopes as a function of neutron number.}
\label{fig:Qalpha_DRHBc_Qalpha_LDM6} 
\end{figure}

The nucleus exhibiting shape coexistence is characterized by the coexistence of different intrinsic shapes at low excitation energy. The onset of shape coexistence is a common phenomenon in superheavy nuclei~\cite{Cwiok:2005nature}, and it is responsible for the observed abrupt shape transitions. The presence of sizable shell gaps in the Nilsson diagram and large barriers between minima on the energy surface are two necessary conditions for nuclei exhibiting shape coexistence. Additionally, intruder states with different parity are important for preventing shape mixing~\cite{Yao:2013PRC}. A quantitative study of the low-energy structure of nuclei with shape coexistence requires beyond mean-field approximation which is beyond the scope of this work.  

Figure~\ref{figs:Nilsson_diagram_Z120}  displays the Nilsson diagram of neutron and proton single-particle energies in $^{304}120$ as a function of the dimensionless quadrupole deformation parameter $\beta^{(t)}_{20}$.  It is seen that for protons, there is a large spherical  shell gap at $Z = 120$, below the $3p_{3/2}$ level, and a relatively smaller spherical shell gap at $Z=114$, formed by the spin-ortbit splitting of the $2f$ states. The evolution of these two shell gaps with neutron number is displayed in Fig.~\ref{fig:spherical_shell_structure_evolution_N}(a). It is shown that the shell gap at $Z=120$ decreases globally with neutron number due to the intruder orbital $1i_{11/2}$. This shell gap is labeled as $\delta_\epsilon$ and displayed quantitatively in Fig.~\ref{fig:two-proton-shell-gaps_Z120}. For comparison, we also plot the two-proton shell gap $\Delta S^{(\rm sph.)}_{2p} $ which is defined as the derivative of the two-proton separation energy,
\beqn
    \Delta S^{(\rm sph.)}_{2p}(A, Z) 
    &=&S^{(\rm sph.)}_{2p}(A,Z) - S^{(\rm sph.)}_{2p}(A+2, Z+2) \nonumber\\
    &=&2B^{(\rm sph.)}(A, Z)-B^{(\rm sph.)}(A-2,Z-2) \nonumber\\
    &&-B^{(\rm sph.)}(A+2, Z+2),
\eeqn
where $B^{(\rm sph.)}(A, Z)$ is the binding energy of the nucleus $(A, Z)$ at the spherical shape from the DRHBc calculation. The two-proton shell gap reflects the underlying nuclear shell structure in the assumption that  there is no dramatic rearrangements of the mean field between the three adjacent nuclei. Indeed, one can see that the evolution trends of the two quantities $\delta_\epsilon$ and $\Delta^{(\rm sph.)}_{2p}$ for $Z=120$ are similar, even though they are quantitatively different.   Moreover, large spherical shell gaps are also found at $Z=92$ and $138$ in Fig.~\ref{fig:spherical_shell_structure_evolution_N}. The former seems to be  in contradiction with the indication from the data on the proton separation energy and $\alpha$-decay width~\cite{Sun:2017PLB}. In contrast, the proton shell gap $Z=126$ is generally small in all the isotopes. 

In Fig.~\ref{fig:spherical_shell_structure_evolution_N}(a), we observe that the size of the $Z=120$ shell gap in light isotopes and the $Z=138$ gap in neutron-rich isotopes are influenced by the splitting of the proton $3p$ states. Smaller splittings correspond to larger shell gaps. The changes of the energy splittings $\Delta \epsilon_{nlj}$ of the proton $3p, 2f$ and $1i$ states as a function of neutron number is displayed in  Fig.~\ref{fig:spin_orbital_splitting_N},  where the energy splitting of spin-orbit doublet states is defined as
\begin{equation}
    \Delta \epsilon_{nlj}=\frac{\epsilon_{nlj_{<}}-\epsilon_{nlj_{>}}}{2l+1},\quad j_{\gtrless}=l\pm1/2.
\end{equation}
It is demonstrated that the energy splitting $\Delta \epsilon_{nlj}$ of the $3p$ doublet states is generally small and can even change sign in the $Z=120$ isotopes with $N=250, 252$, attributed to the formation of evident central depression (also known as bubble structure) in their proton densities, as depicted in Fig.\ref{fig:density_distribution_proton}. The formation of bubble structure induces a significant spin-orbit potential around the nuclear center, which cancels out the contribution around the nuclear surface \cite{Bender:1999PRC} and mainly affects the spin-orbit splitting of low orbital momentum states. These bubble structures also develop in the light $Z=120$ isotopes with $N=172\text{--}182$, explaining the observed weak spin-orbit splittings of the proton $3p$ states in Fig.~\ref{fig:spin_orbital_splitting_N}.

It is also shown in Fig.~\ref{figs:Nilsson_diagram_Z120} that there are two large spherical neutron shell gaps at $N=172$ and $N=184$ in $^{304}120$. However, with the increase of neutron number, the $N=184$ shell gap decreases significantly, while $N=172$ shell gap is rather robust, as shown in Fig.~\ref{fig:spherical_shell_structure_evolution_N}. Moreover, the $N=172$ shell gap  extends from sphericity to an oblate shape with $\beta_{20}$ values down to $-0.15$, explaining the finding in Fig.~\ref{figs:beta2_comparison} that the isotopes with neutron number around 172 are spherical or weakly deformed. On the prolate side, it is seen from Fig.~\ref{figs:Nilsson_diagram_Z120} that there is a large neutron $N=184$ 
 shell gap at $\beta\simeq0.5$, where one also finds a large proton $Z=120$ shell gap. It explains the onset of large prolate deformation in the ground state of $^{304}120$. With the increase of neutron number up to $N=196$, the neutron shell gap shows up around $\beta_{20}=0.4$, cf. Fig.~\ref{figs:Nilsson_diagram_Z120}, explaining the decrease trend of the quadrupole deformation in the $Z=120$ isotopes from $^{304}120$ to $^{316}120$ in Fig.~\ref{figs:beta2_comparison}. In the meantime, a large shell gap around $N=194$ is developed in the oblate side with $\beta_{20}\simeq -0.4$, cf. Fig.~\ref{figs:Nilsson_diagram_Z120}. This explains the development of competing prolate and oblate deformed energy minimum in the isotopes around $N=194$. 
 With the neutron number increases further up to $N=258$, a large spherical neutron shell gap shows up, cf. Fig.~\ref{fig:spherical_shell_structure_evolution_N}.  Consequently, the isotopes with $N\simeq258$ become spherical again. In short, our results indicate that $N=172, 258$ are the next two magic numbers for neutrons in superheavy nuclei beyond $N=126$. This conclusion can also be drawn from the systematic behavior of two-neutron separation energy $S_{2n}(A, Z)$, which  is defined as
\begin{equation}
    S_{2n}(A, Z)=B(A, Z) - B (A-2, Z).
\end{equation}
Here, $B(A, Z)$ is the binding energy of the nucleus $(A, Z)$ from the DRHBc calculation. With the separation energies of two neighbouring nuclei, one can define the two-neutron shell gap
\begin{equation}
    \Delta S_{2n}(A, Z)= S_{2n}(A, Z)-S_{2n}(A+2, Z).
\end{equation}

Figure~\ref{figs:s2ndelta2n} shows the evolution of $S_{2n}$ and $\Delta S_{2n}$ in the four isotopic chains as a function of neutron number. An abrupt drop 
of $S_{2n}(A, Z)$ is found at $N=172$ in the $Z=119, 120$ isotopic chains, at $N=184$ in the $Z=117, 118$ isotopic chains, and at $N=258$ in all the four isotopic chains. These drops correspond to the peaks in the two-neutron shell gap $\Delta S_{2n}(A, Z)$. As shown Fig.~\ref{fig:spherical_shell_structure_evolution_N}, the spherical neutron shell gap $N=258$ decreases  monotonically with neutron number due to intruder state $1k_{15/2}$. In addition, we note that  the size of the $N=172$ shell gap and $N=258$ shell gap are determined by the energy splitting of the neutron $3d$ and $4p$ states, respectively, which are generally small, as shown in Fig.~\ref{fig:spin_orbital_splitting_N}(b). This is also related to the formation of evident bubble structure in corresponding nuclei, as shown in Fig.~\ref{fig:density_distribution_neutron}.  Moreover, it is interesting to see  the sudden increase of the $S_{2n}$ at $N=266$ in Fig.~\ref{figs:s2ndelta2n}, which is attributed to the onset of large prolate deformation, cf.~Fig.~\ref{figs:beta2_comparison}. In contrast, the $S_{2n}$ from the calculation restricted to have the spherical shape decreases smoothly with neutron number, except for $N=258$. Besides, it is shown that in the results of DRHBc calculations, the last bound nuclei of the four isotopic chains are $^{421}117$, $^{424}118$, $^{435}119$, $^{438}120$, respectively. In contrast, the neutron dripline is much more extended if the nuclei are restricted to be spherical in the DRHBc calculation. A similar phenomenon has also been found in Er isotopes~\cite{Wu:2019PRC}.

\subsection{$\alpha$-decay energies}

\begin{figure*}
\includegraphics[width=0.6\paperwidth]{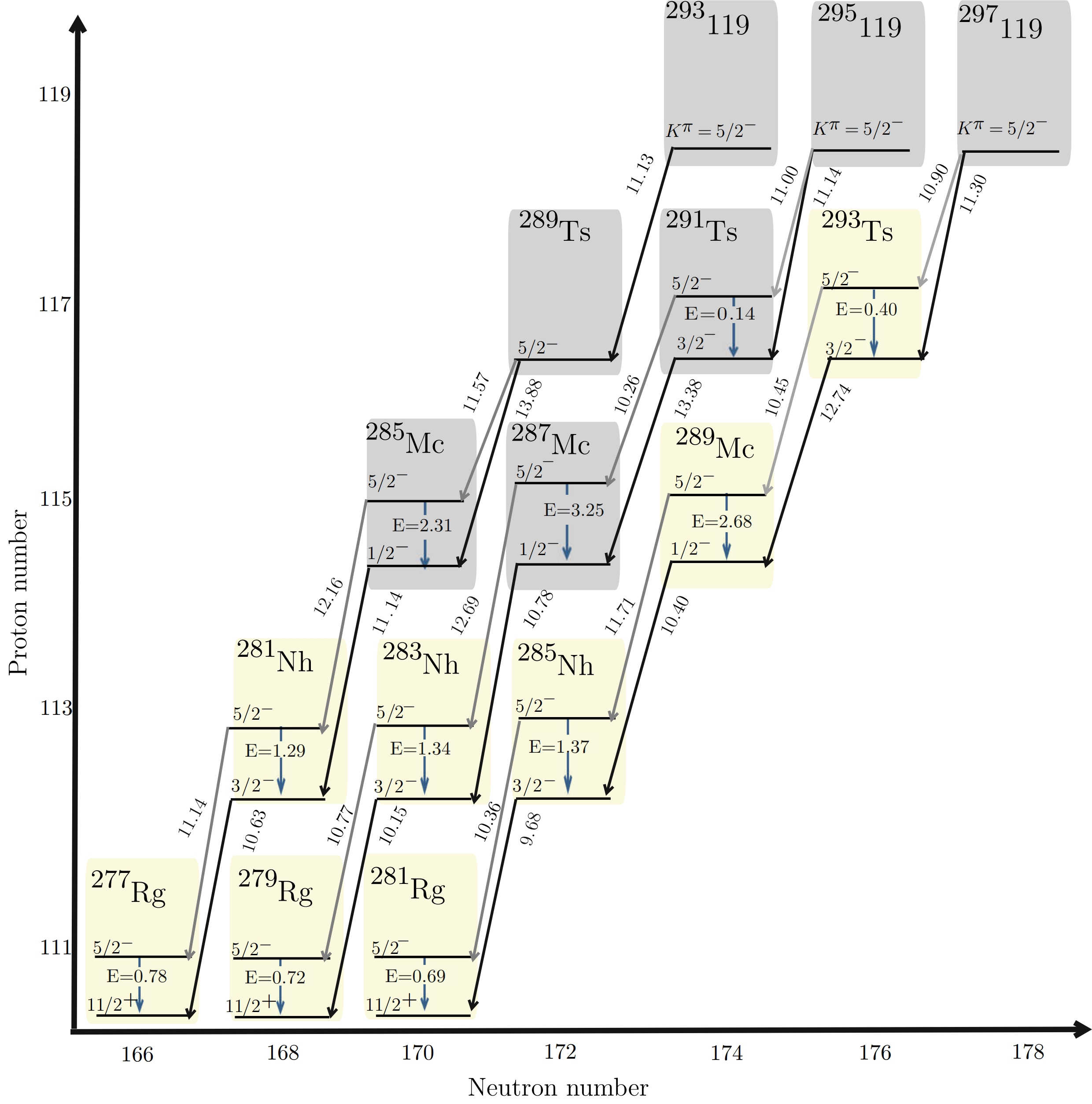}  
	\caption{(Color online) The $\alpha$-decay chains  originating from the ground states of $^{293, 295, 297}$119 from the DRHBc calculation. The lowest-energy state of each nucleus along the chain, together with the state with $K^\pi=5/2^-$, is shown.  The black arrows indicate the favored $\alpha$ decays. The energy difference between two energy levels is given near each arrow. The possible energy levels in between are not concerned. All energies are in MeV. The superheavy nuclei that have already been produced by the \nuclide[249]{Bk}+\nuclide[48]{Ca} reaction~\cite{Oganessian:2011PRC} are indicated with light yellow. See main text for details.}
\label{figs:Qalpha_U1297_Rg281} 
\end{figure*}

The $\alpha$-decay energy $Q_\alpha$ provides a valuable insight into nuclear shell structure~\cite{Heenen:2015fpf}. Figure~\ref{fig:Qalpha}  displays the $Q_\alpha$ values of the four isotopic chains with $Z=117-120$ as a function of neutron number from the DRHBc calculations, where the $\alpha$ decay energies $Q_{\alpha}$ are calculated as
\begin{equation}
    Q_\alpha(A, Z)=B(A-4, Z-2)+B(4,2)-B(A, Z).
\end{equation}
The observed irregular behavior of the $Q_\alpha$ values in each isotopic chain is attributed to shell effects and shape transition. It is shown in Fig.~\ref{fig:Qalpha}(a) that there is a pronounced peak at the neutron number immediately after $N=258$ in the four isotopic chains and an additional peak after $N=184$ in the $Z=117, 118$ isotopes, consistent with the finding in the $Z=108$ isotopes~\cite{Alsultan:2023hxu} . It is also consistent with the behaviors in the two-neutron separations and their differentiates, as seen in Fig.~\ref{figs:s2ndelta2n}, confirming the large shell gaps at $N=258$ and $N=184$. A similar jump at $N=184$ is also observed in the WS4 model for the $Z=117, 118$ isotopes. With the inclusion of rotational correction energies in the DRHBc calculation, one observes more oscillations in the $Q_\alpha$ values in the isotopes around $N=180$. It can be understood from the observation that most of these nuclei are either spherical or weakly deformed. The rotational correction energies are evaluated with the Inglis-Belyaev formula, the validity of which for weakly deformed nuclei and odd-mass nuclei needs to be examined in the future, against the results from the exact angular-momentum projection calculations~\cite{Zhou:2024PRC} based on the same EDF. One may anticipate that the inclusion of beyond mean-field dynamical correlations associated with symmetry restoration and shape mixing will smooth the systematic behavior of the $Q_\alpha$ as a function of the neutron number.  In particular, a new dip shows up at $N=190$. Since there is no data for most of the isotopes, it is difficult to draw a solid conclusion about the underlying shell structure.

Figure~\ref{fig:Qalpha_DRHBc_Qalpha_LDM6} displays the deviation of the $Q_\alpha$ values by the LDM6 model (\ref{eq:LDM6}) from the DRHBc theory for the $Z=119, 120$ isotopes, which is mainly attributed to the shell effects that are missing in the LDM6 model.  In other words, this deviation provides information on the predicted shell structure from the DRHBc theory. Indeed, it is seen that the deviation as a function of neutron number is similar for $Z=119, 120$, where the largest deviations are found around $N=172$ and $N=258$, consistent with the locations of the peaks in the $\Delta S_{2n}$ and $Q_\alpha$ values, cf. Fig.\ref{figs:s2ndelta2n} and Fig.~\ref{fig:Qalpha}, respectively.


It is worth noting that the nonzero angular momenta of the low-lying states of the odd-mass nuclei open more decay channels than those of even-even nuclei. If the angular momenta of the initial and final nuclei during the $\alpha$ decay are different, the emitted $\alpha$ particle will carry a nonzero orbital angular momentum $\ell$. The decay rate of this process will be quenched due to the additional centrifugal barrier between the emitted $\alpha$ and daughter ($D$) nucleus. For the mass number $A\simeq 300$, and distance $R_{\alpha-D}\simeq 10$ fm, the centrifugal barrier is approximately given by
\beq 
V_\ell(R_{\alpha-D})=\frac{\ell(\ell+1)\hbar^2}{2\mu R^2_{\alpha-D}}
\simeq \frac{\ell (\ell+1)}{20}\quad ({\rm MeV}).
\eeq 
According to the estimated formula above, a change of angular momentum by $1\hbar$ in the $\alpha$ decay increases the barrier height by about 0.1 MeV, which may be comparable to the excitation energies of low-lying states in certain daughter nuclei. See, for instance, the excitation energy of $5/2^-$ state in \nuclide[291, 293]{Ts} in Fig.~\ref{figs:Qalpha_U1297_Rg281}.  In such cases, various decay channels may compete with each other. Achieving a quantitative analysis of the branch ratios for decaying into different channels necessitates precise knowledge of the low-lying states of odd-mass nuclei. Describing these states accurately requires going beyond the mean-field approximation~\cite{Zhou:2024PRC}. For simplicity, the energies of states with $J^\pi=K^\pi$ are taken as the energies of the energy-minimal states on the energy curves from the DRHBc calculation. Figure~\ref{figs:Qalpha_U1297_Rg281} illustrates the $\alpha$-decay chain of $^{293, 295, 297}{119}$, which could be produced in the reactions $^{54}$Cr+$^{243}$Am, $^{51}$V+$^{248}$Cm, and $^{50}$Ti+$^{249}$Bk~\cite{Zhu:2023}. According to the DRHBc calculation, the ground state of $^{297}{119}$ has $K^\pi=5/2^-$. Thus, we plot all the states of the nuclei along the $\alpha$-decay chain with $K^\pi=5/2^-$. Moreover, the ground state of each nucleus is also provided. It can be observed that the ground state of \nuclide[293]{Ts} has $K^\pi=3/2^-$, different from that of \nuclide[297]{119}. If the angular momentum of the ground state has the same value as the $K$ value, the relative angular momentum of the $\alpha$ to the daughter nucleus \nuclide[293]{Ts} will be $1\hbar$ for the ground-state to ground-state $\alpha$ decay of \nuclide[297]{119}. Compared to the ground-state to the excited state with $K^\pi=5/2^-$, the centrifugal barrier increases by 0.1 MeV, which is much smaller than the excitation energy (0.45 MeV) of the $5/2^-$ state in \nuclide[293]{Ts}. It indicates that \nuclide[297]{119} will still be dominated by the ground-state to ground-state $\alpha$ decay, even though their ground states have different angular momenta. A similar conclusion also applies to other nuclei along the decay chain of \nuclide[297]{119}. In short, the $\alpha$ decay of odd-mass nuclei is generally much more complicated than that of even-even nuclei, particularly in the presence of rapid shape transitions along the decay chain.

\section{Summary}
\label{Sec.summary}
In this study, we present a comprehensive investigation of neutron-rich odd-$Z$ superheavy nuclei with proton numbers $Z=117$ and $119$, and neutron numbers $N$ ranging from $N=170$ to the neutron dripline. We employ the axially deformed relativistic Hartree-Bogoliubov theory in continuum, which has achieved significant success in the global study of atomic nuclei across the nuclear chart. The treatment of unpaired valence nucleons utilizes the "automatic blocking" procedure under the equal filling approximation. The ground state is determined as the state with the lowest energy after the self-consistent calculation, even though several states with competing energies may coexist in some nuclei. The obtained results are compared to neighboring isotopic chains with $Z=118$ and $120$.

We extensively explore the evolution of shell structure and shape transition in the four isotopic chains at the mean-field level. The examination of shape evolution is based on the bulk properties of the energy-minimal states on the potential-energy curves. Our investigation reveals that the ground states of the majority of these superheavy nuclei are prolate deformed, except for nuclei with proton numbers $Z=119, 120$, and neutron numbers around $N=200$, where nearly-degenerate oblate deformed states are observed. Specifically, we find that the ground states of nuclei in the four isotopic chains are spherical or weakly deformed with neutron numbers $N<186$. Beyond this neutron number, the isotopes exhibit significant prolate or oblate deformation with $\beta_{20}\simeq 0.6$ or $\beta_{20}\simeq -0.4$. This quadrupole deformation gradually decreases to zero as the neutron number increases up to $N=248$. At $N=268$, the ground states of the four isotopic chains become prolate deformed again with $\beta_{20}\simeq 0.4$, which then slowly decreases to 0.2 towards the neutron dripline. This evolution trend with neutron number is also reflected in the systematic behaviors of the neutron radii and charge radii. Comparatively, the low-energy structure of odd-$Z$ superheavy nuclei is shown to be more intricate. This is evidenced by significant changes in the quantum numbers $K^\pi$ of the ground state of odd-mass nuclei as the predominant shape evolves with neutron number. Such variations may lead to changes in angular momentum in $\alpha$ decay, thereby affecting the decay rate.  

The results of the DRHBc calculations, including Nilsson diagrams, quadrupole deformations, rms radii, and $Q_\alpha$ values, suggest that  $Z=120$ is the next proton magic number and $N=172$ as a neutron magic number. Additionally, $N=258$ is identified as a potential neutron magic number in the neutron-rich region. The emergence of these neutron shell gaps is attributed to the presence of nearly-degenerate neutron $3d$ and $4p$ spin-orbit doublet states, facilitated by the bubble structure. It is noteworthy that the potential energy curves of many nuclei near $Z=120$ and $N=184$ display softness, with variations in quadrupole deformation, suggesting potential significant influences of dynamical correlations on the ground-state properties of these superheavy nuclei. The impact of these correlations on nuclear structural properties warrants further investigation in future studies.

\section*{Acknowledgments} 

We thank C. F. Jiao, N. Wang, Y. N. Zhang, and L. Zhu, as well as all members of the DRHBc Mass Table Collaboration, for their fruitful discussions. Special thanks are owed to J.M. Dong, C. Pan, and C. Zhou for their critical reading of the manuscript. This work is partly supported by the National Natural Science Foundation of China (Grant Nos. 12141501, 12275369, and 12305125), Guangdong Basic and Applied Basic Research Foundation (2023A1515010936), the Natural Science Foundation of Sichuan Province (24NSFSC5910), and the Fundamental Research Funds for the Central Universities, Sun Yat-sen University.

\bibliographystyle{apsrev4-1} 

%

\end{document}